\documentclass[%
 reprint,
superscriptaddress,
 amsmath,amssymb,
 aps,
]{revtex4-1}

\usepackage{graphicx}
\usepackage{dcolumn}
\usepackage{bm}
\usepackage[colorlinks= true, linkcolor=blue, citecolor=red, urlcolor=red]{hyperref}
\usepackage{lipsum}
\usepackage{bbold}
\usepackage{mathtools}
\usepackage[normalem]{ulem}
\usepackage{enumitem}
\usepackage[T1]{fontenc}

\def\bea{\begin{eqnarray}}
\def\eea{\end{eqnarray}}

\makeatletter
\newcommand{\vast}{\bBigg@{4}}
\newcommand{\Vast}{\bBigg@{5}}
\makeatother

\newcommand{\aref}[1]{Appendix \ref{#1}}%

\usepackage{xcolor}

\newcommand{\eref}[1]{Eq.~(\ref{#1})}
\newcommand{\fref}[1]{Fig.~\ref{#1}} 
\newcommand{\wt}[1]{\widetilde{#1}}

\begin{document}

\title{A resetting particle embedded in a viscoelastic bath}

\author{Arup Biswas}
\email{arupb@imsc.res.in}
\affiliation{The Institute of Mathematical Sciences, CIT Campus, Taramani, Chennai 600113, India \& Homi Bhabha National Institute, Training School Complex, Anushakti Nagar, Mumbai 400094, India}

\author{Johan L.A. Dubbeldam}
\email{J.L.A.Dubbeldam@tudelft.nl}
\affiliation{Delft Institute of Applied Mathematics, Delft University of Technology, 2628 CD Delft, The Netherlands}

\author{Trifce Sandev}
\email{trifce.sandev@manu.edu.mk}
\affiliation{Research Center for Computer Science and Information Technologies, Macedonian Academy of Sciences and Arts,
Bul. Krste Misirkov 2, 1000 Skopje, Macedonia} \affiliation{Institute of Physics, Faculty of Natural Sciences and Mathematics, Ss. Cyril and Methodius University,
Arhimedova 3, 1000 Skopje, Macedonia} \affiliation{Department of Physics, Korea University, Seoul 02841, Korea}

\author{Arnab Pal}
\email{arnabpal@imsc.res.in}
\affiliation{The Institute of Mathematical Sciences, CIT Campus, Taramani, Chennai 600113, India \& Homi Bhabha National Institute, Training School Complex, Anushakti Nagar, Mumbai 400094, India}

\date{\today}

\begin{abstract}
We examine the behavior of a colloidal particle immersed in a viscoelastic bath undergoing stochastic resetting at a rate $r$. Microscopic probes suspended in viscoelastic environment do not follow the classical theory of Brownian motion. This is primarily because the memory from successive collisions between the medium particles and the probes does not necessarily decay instantly as opposed to the classical Langevin equation. To treat such a system one needs to incorporate the memory effects to the Langevin equation. The resulting equation formulated by Kubo, known as the Generalized Langevin equation (GLE), has been instrumental to describe the transport of particles in inhomogeneous or viscoelastic environments. The purpose of this work, henceforth, is to study the behavior of such a colloidal particle governed by the GLE under resetting dynamics. To this end, we extend the renewal formalism to compute the general expression for the position variance and the correlation function of the resetting particle driven by the environmental memory. These generic results are then illustrated for the prototypical example of the Jeffreys viscoelastic fluid model. In particular, we identify various timescales and intermittent plateaus in the transient phase before the system relaxes to the steady state; and further discuss the effect of resetting pertaining to these behaviors. Our results are supported by numerical simulations showing an excellent agreement.
\end{abstract}

\pacs{Valid PACS appear here}
\maketitle

\begin{quotation}
\noindent
\textbf{Diffusion process of a colloidal particle or a polystyrene bead immersed in a fluid such as water is a cornerstone in statistical physics. If the surrounding fluid molecules are smaller and faster than the probe, a distinct separation of timescales can be observed which results in a dynamics that is Markovian or memoryless in nature. It is well known that such dynamics can be described by the celebrated Langevin equation. Nonetheless, this is no longer the case when the probes are driven through solutions in the presence of long macromolecules, a dense environment or a viscoelastic medium. Such dynamics are more complex, giving rise to memory effects and are quantified by the generalized Langevin equations (GLE). We aim to study the GLE under stochastic resetting dynamics which has emerged as a powerful mechanism to stabilize the system by eliminating the wandering-off or kinetically trapped trajectories. Our analysis reveals that resetting not only induces a stationary state into the system but also allows one to harness the timescales arising from the memory effect. Our research opens door to design resetting based strategies to explore non-equilibrium transport phenomena in complex fluid.}
\end{quotation}

\section{Introduction}
The classical theory of Brownian motion describes, for example, the random motion of a massive particle immersed in a fluid, as it was observed by Robert Brown in 1827 with pollen grains and dust particles in water. The random motion of the particle occurs due to the thermal motion of the molecules in the liquid, mass of the latter is much smaller than the mass of the 
suspended particle. Paul~Langevin explained this motion of the Brownian particle with mass $m$ by the Newton's second law for a test particle in presence of viscous dynamic friction $-\gamma_fv(t)$, deterministic external potential $V(x)$ and an internal random force $\xi(t)$, \textit{i.e.},~\cite{langevin1908theorie}
\begin{align}
    m\dot{v}(t)+\gamma_f v(t)+\frac{dV(x)}{dx}=\xi(t), \quad \dot{x}(t)=v(t), \label{le}
\end{align}
where $x(t)$ and $v(t)$ are the particle displacement and particle velocity, respectively, and $\gamma_f$ is the friction coefficient. The internal force $\xi(t)$ is the Gaussian random force of zero mean ($\langle\xi(t)\rangle=0$) and correlation $\langle\xi(t)\xi(t')\rangle=2\gamma_fk_{B}T\delta(t-t')$, where $\langle\cdot\rangle$ means ensemble average. This means that the noise is internal white noise and the fluctuation and dissipation in the system come from a same source. The time scale of the molecular motion is much shorter than the time scale of the motion of the Brownian particle. The resulting mean squared displacement (MSD) in absence of external potential ($V(x)=0$) in long time limit shows a linear time dependence, $\langle x^{2}(t)\rangle\sim t$, which is characteristic for normal diffusion, while at short time at short times the motion is ballistic due to the inertial effects. The transition from ballistic motion to normal diffusion is characterized by the characteristic time scale $1/\gamma_f$.

In many systems, the mass of the immersed particle is not necessarily much larger than the mass of the surrounding molecules of the environment, and consequently, the time scale of the molecular motion is not very much shorter than the time scale of the motion of the suspended particle. In such cases, the Langevin equation for the Brownian motion should be modified to the generalized Langevin equation with friction memory kernel~\cite{mori1965transport,kubo1966fluctuation,hanggi1978correlation,haunggi1994colored}, which is of interest to our work.  

The generalized Langevin equation has been used to model anomalous diffusion by employing power-law friction memory kernel~\cite{lutz2001fractional,vinales2006anomalous,desposito2008memory,burov2008fractional,desposito2009subdiffusive,sandev2011generalized}. The anomalous diffusion is characterized by power-law dependence of the MSD on time, $\langle x^2(t)\rangle\sim t^{\alpha}$, $\alpha\ne1$, which has been observed in various systems, from electron transfer within a single protein molecule~\cite{yang2003protein,kou2004generalized}, to models of solute particle in a bath of fast solvent molecules~\cite{kneller2014communication} and particles in viscoelastic media~\cite{goychuk2012viscoelastic}. The generalized Langevin equation model has been used, as well as, in description of the conformational motions of proteins~\cite{lange2006collective,lee2019multi}, in the microscopic description of a tracer particle in a one-dimensional many-particle system with two-body interaction potential~\cite{lizana2010foundation}, in generalized elastic model of stochastic motion in membranes and semiflexible polymers~\cite{taloni2010generalized} and polymer translocation~\cite{jlad}, to mention but a few. Here we mention that a simple example of emergence of exponential memory is in the model of Brownian motion in presence of a harmonic potential, \textit{i.e.}, in the Ornstein-Uhlenbeck (OU) process~\cite{zwanzig2001nonequilibrium,wang1996linear,adelman1976fokker}. In this work, we will be interested in the Jeffreys fluid model for a particle in a viscoelastic medium which is also an example of non-Markovian system ~\cite{ferrer2021fluid,gomez2015transient,das2023enhanced,paul2021bayesian,gomez2016dynamics,gomez2021work,darabi2023stochastic,paul2018free,straube2024memory}. In particular, the goal is to monitor the motion of a particle in such fluid system under the resetting dynamics as will be delineated in below. 

Contrary to the exhaustive analysis of various diffusion and anomalous processes governed by different (generalized) diffusion and Fokker-Planck equations for homogeneous and heterogeneous media in presence of stochastic resetting~\cite{evans2011diffusion,evans2020stochastic,pal2015diffusion,sandev2022heterogeneous,wang2021time,sandev2022stochastic,lenzi2022transient,kusmierz2019subdiffusive,mendez2022nonstandard,mendez2021continuous,pal2023random}, there is a lack of works related to the generalized Langevin equation in presence of stochastic resetting. In resetting processes, a moving particle is reset to its initial (or particular) location at regular or stochastic intervals. Quite interestingly, resetting has the ability to stabilize a system by repeatedly reverting it to a fixed location. This was first observed by Evans and Majumdar in simple diffusion which, in the absence of resetting is a non-stationary process, but attains stationarity as soon as  resetting is introduced \cite{evans2011diffusion}.  
Similar effect has also been observed in the above mentioned models of diffusion and Fokker-Planck equations in the presence of resetting where the system approaches a non-equilibrium stationary state (NESS)~\cite{evans2011diffusion,evans2020stochastic,pal2023random,mendez2016characterization,stojkoski2021geometric,eule2016non,tal2022diffusion,ray2020diffusion}, and the relaxation dynamics to the stationary state is shown to be far from trivial~\cite{majumdar2015dynamical,evans2020stochastic,singh2020resetting}.  

\begin{figure}
    \centering
\includegraphics[width=8cm]{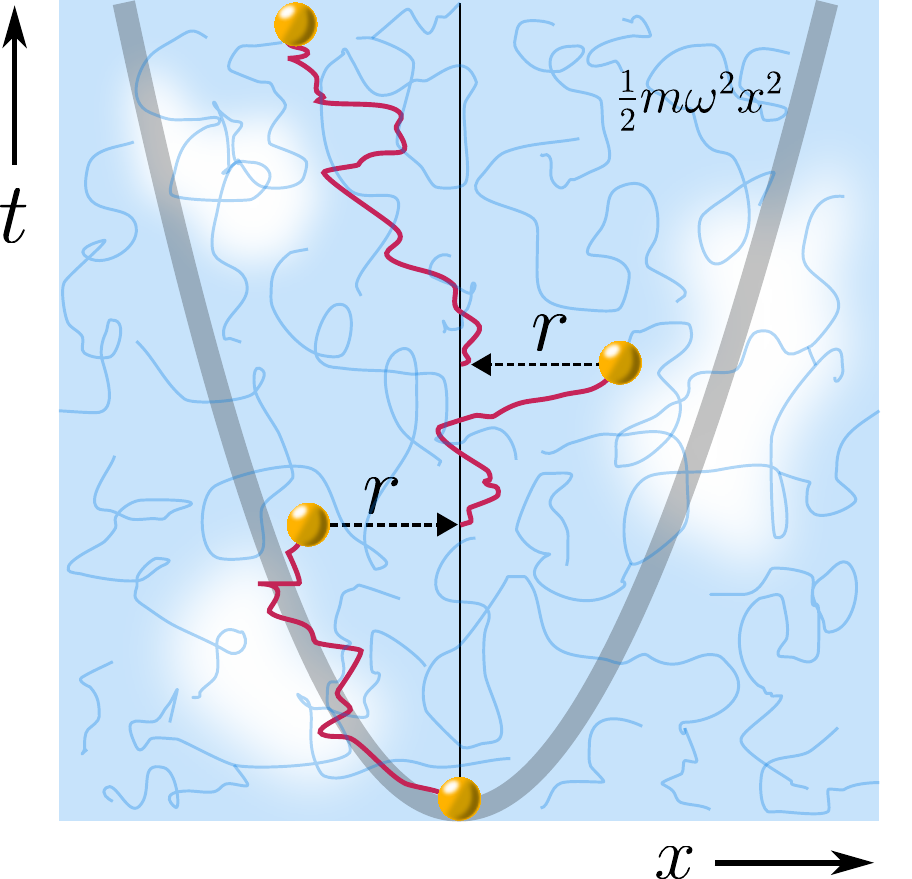}
    \caption{Schematic representation of a colloidal particle diffusing in a viscoelastic bath under harmonic trapping. The motion of the particle is governed by the generalized Langevin equation (GLE) as in \eref{gle-ovd}. In addition, the particle undergoes stochastic resetting at random time intervals drawn from an exponential distribution with mean $1/r$.}
    \label{fig0}
\end{figure}

In the present work, we consider a process described by the generalized Langevin equation in presence of Poissonian resetting. This means that the process is renewed at random times that follow an exponential distribution given by $f_{R}(t)=re^{-rt}$, where $1/r$ is mean resetting time (ie, $r$ is the resetting rate). We give a detailed analysis of the MSD and correlation functions for the general form of the friction memory kernel. Then, we apply our results to a particular form of the friction memory kernel, which is used in the Jeffreys fluid model. We confirm our analytical findings by numerical simulations. 

The paper is organized as follows: At first (section \ref{secI}) we discuss in detail about the generalized Langevin equation (GLE) and solve them to find the MSD and correlation function of a particle in a viscoelastic media with arbitrary kernel. Then we elaborate on the renewal formalism of stochastic resetting in section \ref{secIII} to find the general expression for the mean and correlation function of a particle following GLE and subjected to stochastic resetting at rate $r$. In section \ref{secIV} we take the example of Jeffrey's fluid to model a viscoelastic bath to illustrate in detail about the behaviour of mean and correlation function both with and without resetting. We verify our analytical findings with numerical simulation technique, namely the Markovian embedding scheme, as discussed in section \ref{sec5}. Finally, we conclude with a brief summary of the work and future outlook in section \ref{secVI}.

\section{Generalized Langevin equation} \label{secI}
Let us consider a particle of mass $m$ in a viscoelastic bath with correlated thermal noise $\xi(t)$. Moreover, the particle is placed under a potential  $V(x)$. The generalized Langevin equation (GLE) for the particle then can be written as~\cite{kubo1966fluctuation,zwanzig2001nonequilibrium}
\begin{align}
\label{GLE}
m\ddot{x}(t)+m\int_0^t\gamma(t-t')\dot{x}(t')dt'+\frac{d
V(x)}{dx}&=\xi(t).
\end{align}
The quantity $\gamma(t)$ is the generalized friction kernel which correlates the velocity of the particle at different times. In other words, this can be understood as the non-Markovian response of the particles in the fluid. The generalized fluctuation-dissipation theorem for GLE case is given by~\cite{kubo1966fluctuation,zwanzig2001nonequilibrium}
\begin{align}
    \langle \xi(t)\xi(t') \rangle=k_BTm\gamma(|t-t'|). \label{fdt}
\end{align}
Evidently when $\gamma(t-t')=2\gamma_f\delta(t-t')$ we recover the standard Langevin equation with a constant friction $\gamma_f$ (note that, the factor 2 is canceled out from the integration in Eq. (1) as the upper limit of the integration runs upto $t$ which is the same point where the delta function is infinity). The correlation, as a consequence, also takes the familiar form $\langle \xi(t)\xi(t') \rangle=k_BTm\gamma_f\delta(t-t')$. We assume the external potential $V(x)$ to be harmonic in nature so that 
\begin{align}
    V(x)=\frac{1}{2}m\omega^2x^2.
\end{align}

For the experimental conditions, in many systems, the inertial term $m\ddot{x}(t)$ in the underdamped GLE \eref{GLE} can be neglected, which means that the friction in the system is very large. This motion is known as overdamped motion~\cite{kampen}. Such case of high viscous damping (large friction) is considered to model experimental data which are related, for example, to the movement of driven colloids in aqueous solution of poly-ethylene oxide which is a polymer that provides elasticity
along with its inherent viscosity and renders the  viscoelastic solution damped \cite{das2023enhanced,paul2021bayesian}. Similarly, due to the liquid environment of proteins, frictional term is usually very high, and thus the motion of the macromolecules could be considered overdamped, see for example Refs.~\cite{10.1214/07-AOAS149,min2005observation}. Moreover, the movement within proteins is confined to a short range, and the potential can be well approximated by a harmonic potential, making the overdamped GLE for a harmonic oscillator a suitable model for description of the dynamics within proteins~\cite{10.1214/07-AOAS149,min2005observation,kou2004generalized}.

Under these assumptions, the resulting GLE in the overdamped limit takes the form
\begin{align}
m\int_0^t\gamma(t-t')\dot{x}(t')dt'+m\omega^2x=\xi(t). \label{gle-ovd}
\end{align}
Our aim is to analyze this overdamped GLE in the presence of stochastic resetting. However, it is instructive to first revisit the techniques and solutions of GLE in the absence of resetting and find the relevant quantities of our interest. Building upon these results, we will develop methods when resetting is introduced with further elucidation of the key results.

\subsection*{\textbf{\textit{GLE in Laplace space: Correlation function and MSD}}}
We start by taking Laplace transform on both sides of  \eref{gle-ovd} and dividing by $m$, this yields
\begin{align}
   \widetilde{\gamma}(s)(s\widetilde{X}(s)-x_0)+\omega^2\widetilde{X}(s)=\frac{\widetilde{\Xi}(s)}{m},
\end{align}
where we denote the Laplace transformed quantities $\widetilde{X}(s)=\int_0^{\infty}dt~e^{-st}x(t),~\widetilde{\Xi}(s)=\int_0^{\infty}dt~e^{-st}\xi(t),~\widetilde{\gamma}(s)=\int_0^{\infty}dt~e^{-st}\gamma(t)$ and $x_0=x(t=0)$.
After a slight rearrangement one obtains
\begin{align}
    &\wt X(s)= \left[ \frac{1}{s}- \omega^2 \wt I_0(s) \right] x_0+ \frac{1}{m}\wt \Xi(s) \wt G_0(s), \label{xs-ovd}
\end{align}
where the functions $G_0(t)$ and $ I_0(t)$ are the so-called relaxation functions defined as
\begin{align}
    &G_0(t)=\mathcal{L}^{-1}\left[\wt G_0(s)\right]=\mathcal{L}^{-1}\left[\frac{1}{s\widetilde{\gamma}(s)+\omega^2}\right], \label{g0}\\ 
    &I_0(t)=\mathcal{L}^{-1}\left[\wt I_0(s)\right]=\mathcal{L}^{-1}\left[\frac{s^{-1}}{s\widetilde{\gamma}(s)+\omega^2}\right]. \label{i0}
\end{align}
Here the operator $\mathcal{L}^{-1}[\wt f(s)]$ stands for the Laplace inversion of the function $\wt f(s)$.
\eref{xs-ovd} can be inverted to write an integral solution for $x(t)$ as follows
\begin{align}
 x(t)= \langle x(t) \rangle + \frac{1}{m}\int_0^t dt'~G(t-t')\xi(t'),
\end{align}
where $\langle x(t) \rangle=\left[ 1-\omega^2 I_0(t) \right] x_0$. In what follows we will assume $x_0=0$ (unless stated otherwise) without loss of any generality. From \eref{xs-ovd}, 
the correlation function in the Laplace domain is then given by
\begin{align}
    \langle \wt X(s) \wt X(s') \rangle=\frac{1}{m^2}  \wt G_0(s)\wt G_0(s')\langle\wt \Xi(s)\wt \Xi(s') \rangle,
\end{align}
where
the quantity $\langle\wt \Xi(s)\wt \Xi(s') \rangle$ in the above equation can be computed from \eref{fdt} by performing a double Laplace transform \cite{pottier2003aging} as
\begin{align}
    \langle \wt\Xi(s)\wt\Xi(s') \rangle=k_BTm \left(\frac{\wt \gamma(s)+\wt \gamma (s')}{s+s'}\right). \label{xis}
\end{align}
Writing $\wt \gamma(s)$ in terms of the relaxation functions using \eref{g0}-(\ref{i0}) and after a bit of simplifications, we finally arrive at
\begin{align}
    \langle \wt X(s) \wt X(s') \rangle&=\frac{1}{m^2}\wt G_0(s)\wt G_0(s')\langle \wt \Xi(s)\wt\Xi(s') \rangle \nonumber\\
    &=\frac{k_B T}{m}\Bigg[\frac{\wt I_0(s)}{s'}+\frac{\wt I_0(s')}{s}\nonumber\\& \hspace{0.5cm}-\frac{\wt I_0(s)+\wt I_0(s')}{s+s'}-\omega^2\wt I_0(s)\wt I_0(s')\Bigg]. \label{gsgs-ovd}
\end{align}
The resulting correlation in the time domain takes the form
\begin{align}
  &C(t,t')=\langle x(t) x(t') \rangle\nonumber\\
  &=\frac{k_B T}{m}\big[I_0(t)+I_0(t')-I_0(|t-t'|)-\omega^2I_0(t)I_0(t')\big]. \label{xclt-ovd}
\end{align}
The MSD can be obtained by setting $t=t'$ in the correlation function (assuming $x_0=0$ so that $\langle x(t) \rangle=0$) which is given by
\begin{align}
      &\langle x^2(t) \rangle=\frac{k_B T}{m}\left[2I_0(t)-\omega^2I^2_0(t)\right].\label{x2t}
\end{align}
Note that we dropped the term $I_0(0)$ since from \eref{i0} and according to the initial value theorem $I_0(t=0)=\lim_{s\rightarrow\infty}s\tilde{I}_{0}(s)=0$. With a specific choice of the friction kernel one can explicitly evaluate the correlation function of a particle following GLE. We shall elaborate on the same for the Jeffreys fluid model in the later part of this article. In the next section, we introduce resetting to the GLE and elaborate on the theory to derive exact expressions for the MSD and the correlation function.

\section{GLE with stochastic resetting} \label{secIII}
Under resetting dynamics, motion of a particle, that follows a GLE, is intermittently stopped and the particle is brought back to the origin $x=0$. This dynamics repeats itself at stochastic time intervals where the times are drawn from an exponential distribution $f_R(t)=re^{-r t}$ with mean $1/r$. Between two consecutive resetting events, the particle follows the GLE as in \eref{gle-ovd} initiating from the same coordinate each time. Let us denote $P(x,t)$ as the probability density of the particle to be found at $x$ at time $t$, starting from $x_0=0$ at $t=0$, in the absence of resetting. Using renewal techniques \cite{evans2014diffusion,maso2019transport,evans2020stochastic,pal2023random}, one then can write the corresponding propagator $P_r(x,t)$ for the same system undergoing resetting as 
\begin{align}
    P_r(x,t)=e^{-rt}P(x,t)+r\int_0^t d\tau e^{-r \tau}P(x,\tau). \label{ren}
\end{align}
The physical interpretation of the above equation is as follows: The first term accounts for those trajectories that did not undergo any resetting event up to time $t$ which occurs with a small probability $e^{-rt}$ multiplied by the reset-free propagator $P(x,t)$. The second term on the other hand takes into account for all the possible trajectories which have encountered at least one resetting event. In particular, we assume the last resetting event to occur at time $t-\tau$. The probability that a resetting event occurs between time $\tau$ and $\tau+d\tau$ is $rd\tau$ which is multiplied with the probability that no resetting event occurred after $t-\tau$ which is $e^{-r \tau}$. After the last resetting event at $t=\tau$ the dynamics follows the reset-free propagator $P(x,t)$ for the remaining time $\tau$. 

A subtle point to note here is that the renewal equation \eref{ren} holds only when the full dynamics (not only the position) of the particle is reset. In our case, the friction kernel $\gamma(t)$ is a time-dependent quantity. Thus, at each resetting event, we need to restart $\gamma(t)$ to its starting value as well. That is, the memory kernel keeps the dynamics non-Markovian between the resetting intervals while the resetting event is simply Markovian. This allows us to take advantage of the \textit{full renewal formalism}. This argument is also illustrated in section \ref{sec5} where we discuss the simulation schemes. Let us now find the exact form of the MSD and correlation function discussed earlier in presence of stochastic resetting.

\subsection{MSD in presence of resetting}
We can use \eref{ren} to obtain a renewal equation for the MSD under resetting \cite{mendez2016characterization,stojkoski2021geometric} 
\begin{align}
  \langle x^2 (t) \rangle_r&=  \int_{-\infty}^{\infty}dx~x^2P_r(x,t) \nonumber\\
  &=e^{-rt}\langle x^2 (t) \rangle+r\int_0^t d\tau e^{-r \tau}\langle x^2 (t) \rangle,  \label{x2r}
\end{align}
where $\langle x^2 (t) \rangle= \int_{-\infty}^{\infty}dx~x^2P(x,t)$ is the MSD of the underlying process which is given in \eref{x2t}. Finally plugging the result from \eref{x2t} in \eref{x2r} we obtain the exact formula for MSD with resetting given as
\begin{align}
   \langle x^2(t) \rangle_r=&\left(\frac{k_B T}{m}\right)\Big(e^{-rt}\left[2 I_0(t)-\omega^2 I_0^2(t)\right] \nonumber\\&+ r\int_0^t d\tau e^{-r \tau}\left[2 I_0(\tau)-\omega^2 I_0^2(\tau)\right]\Big).\label{msd-r}
\end{align} 
The above equation is general and it holds for any friction kernel. Since the system reaches a non-equilibrium steady state (NESS) under resetting in the long time limit, this is reflected in the MSD as well. 
To see this, we set $t \to \infty$ in \eref{msd-r}. There, the first term vanishes in the long time limit and the MSD in the NESS for GLE reads
\begin{align}
    \langle x^2 \rangle_r^{ss}&=\langle x^2(t\to \infty)\rangle_r =\left(\frac{k_B T r}{m}\right)\Big[2\wt I_0(r)-\omega^2 \widetilde{I_0^2}(r)\Big],
\end{align}
where recall that $\wt I_0(r)$ is the Laplace transform of $I_0(t)$ and $\widetilde{I_0^2}(r)$ denotes the Laplace transform of $I_0^2(t)$ with variable $r$. Besides the MSD, one can also try to extract information about the autocorrelation function of the GLE under resetting. For this purpose, one needs to write a separate renewal equation for the autocorrelation function itself. In the next subsection, we proceed to illustrate the same.

\subsection{Correlation function with resetting}
The autocorrelation function for resetting systems have been studied recently in the context of diffusion \cite{stojkoski2022autocorrelation} and fractional Brownian motion \cite{majumdar2018spectral}. In here, we adapt the renewal structure that was proposed in \cite{majumdar2018spectral}. For brevity, we briefly revisit the derivation in below. The autocorrelation function of 
 $x(t)$ under resetting dynamics is defined in the following way: $C_r(t,t') = \langle x(t) x(t') \rangle_r$ which satisfies the following renewal relation for $t> t'$,
\begin{align}
    C_r(t,t')=&e^{-rt}C(t,t')\nonumber\\
    &+r\int_0^{t'} d\tau e^{-r(t-t'+\tau)}C(\tau,t-t'+\tau) . \label{corr-reset}
\end{align}
First, note that to estimate a meaningful correlation between time $t'$ and $t$ one needs to ensure that no resetting event has taken place in that selected time interval. This is because any resetting in the same interval would reset the particle to the origin washing out any correlation between the particle's position at time $t'$ and the same at a later time $t$. 

The renewal structure of Eq. \eqref{corr-reset} can be interpreted as follows. The first term on the RHS accounts for those trajectories which did not undergo any resetting event for the entire observation window $[0,t]$ -- this survival probability is given by $e^{-rt}$ and then it should be multiplied with $C(t,t')$ \textit{i.e.}, the correlation function of the underlying process. The other possibility is to have multiple resetting events prior to the interval $[t',t]$. Let us assume that the last resetting event had happened $\tau$ amount of time before $t'$ \textit{i.e.}, at the time instant $t'-\tau$. Starting from here the motion of the particle renews. As a consequence, the correlation between $t'$ and $t$ is effectively the same as the correlation between $\tau$ and $t- (t'-\tau)$ measured with respect to $t'-\tau$ as the new time origin. After $t'-\tau$ no resetting event takes place for the remaining time $t-t'+\tau$ with probability $e^{-(t-t'+\tau)}$. Moreover, the probability that a resetting event occurs between time $t'-\tau$ and $t'-\tau+d\tau$ is simply $rd\tau$. Both these factors are multiplied to the underlying process's correlation function between $\tau$ and $t-t'+\tau$ where no resetting event occurs. Since $\tau$ can occur any time between 
$0$ to $t'$, we integrate this time out to arrive at the second term in 
 \eref{corr-reset}.

The relation \eref{corr-reset} is quite useful since by simply plugging the correlation function of the underlying process as in \eref{xclt-ovd}, one can obtain the same under resetting. To proceed to the exact evaluation of MSD (\eref{msd-r}) and correlation function (\eref{corr-reset}), one requires specific choice of the kernel $\gamma(t)$ (and hence the relaxation functions). In what follows, we illustrate our general results by taking an example of the Jeffreys fluid model, which has been a paradigmatic choice to model viscoelastic systems.

\section{Jeffreys fluid as a viscoelastic bath} \label{secIV}
Jeffreys fluid model has been found to be a good representation of a viscoelastic bath in several experimental systems \cite{ferrer2021fluid,gomez2015transient,raikher2013brownian,das2023enhanced,paul2021bayesian,gomez2021work}. It is also an intuitive yet illustrative model that can capture both the viscous and elastic timescales of the bath. The viscous part is given by the delta-correlated kernel and the elastic part is given by a mono-exponential function. This makes the model tractable analytically and by numerical simulations. 

The friction kernel $\gamma(t)$ in the  Jeffreys fluid model is usually considered in the following manner \cite{ferrer2021fluid,gomez2015transient,raikher2013brownian,das2023enhanced,paul2021bayesian,gomez2021work}
\begin{align}
    &\gamma(t)=2 \gamma_f \delta(t)+\frac{\gamma_s}{\tau_s} \exp \left( -\frac{t}{\tau_s}  \right). \label{jfm-1}
\end{align}
Here the parameter $\gamma_f$ relates the viscous property of the bath to the particle under consideration. The other parameters in the exponential \textit{i.e.} $\gamma_s$ and $\tau_s$ determine the elastic properties of the fluid. A larger value of $\tau_s$ implies that the particles of the fluid relax very slowly whereas $\gamma_s$ measures the strength of the relaxation dynamics of the fluid in the particle's dynamics. In the limit of $\gamma_s=0$ one recovers the result for usual Langevin dynamics with diffusion constant $D=\frac{k_B T}{m \gamma_f}$.

To move forward, we take Laplace transform of \eref{jfm-1} resulting in $\wt \gamma(s)=\gamma_f +\frac{\gamma_s}{1+s\tau_s}$. 
Using the above in \eref{i0} we find
\begin{align}
    &\wt I_0(s)=\frac{s^{-1}}{s\left(\gamma_f +\frac{\gamma_s}{1+s\tau_s}\right)+\omega^2},
    \end{align}
which upon Laplace inversion yields
\begin{align}
  &I_0(t)=\frac{1}{\omega^2}\Bigg(1-e^{-\alpha t/\tau_s}\cosh{\left[\frac{t}{\tau_s}\sqrt{\alpha^2-\beta^2}\right]}\nonumber\\
    &\hspace{1cm}-\frac{\alpha-\beta^2}{\sqrt{\alpha^2-\beta^2}}e^{-\alpha t/\tau_s}\sinh{\left[\frac{t}{\tau_s}\sqrt{\alpha^2-\beta^2}\right]}\Bigg), \label{i0t-jeff}
\end{align}
where we have defined the following dimensionless quantities 
\begin{align}
    \alpha=\frac{\gamma_f+\gamma_s+\tau_s\omega^2}{2\gamma_f}, ~~ \beta=\sqrt{\frac{\omega^2\tau_s}{\gamma_f}}. \label{al-be}
\end{align}
It is now straightforward to plug this expression in \eref{msd-r} and \eref{corr-reset} to get the results for MSD and correlation function with resetting, respectively.

\subsection{MSD}

In the following, we elaborate in detail about behavior and different timescales arising in the expression of MSD in both the reset-free and resetting induced process. Let us start with the reset-free process first.
\begin{figure}
    \centering
    \includegraphics[width=8.5cm]{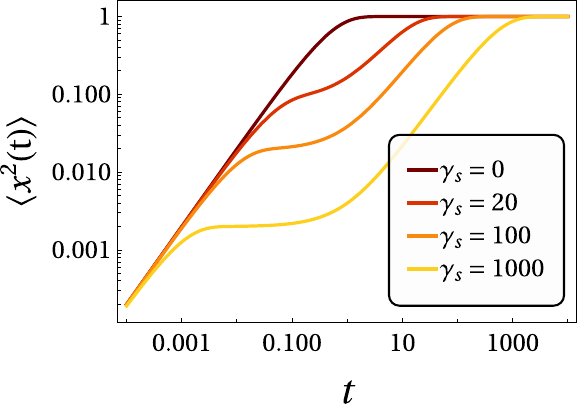}
    \caption{MSD of the underlying reset-free process for the Jeffreys fluid model as a function of time for different values of $\gamma_s$. The circles represent results from the simulation. For $\gamma_s=0$, the MSD is the same as that of an OU process.  The plots indicate the existence of two distinct time scales for a non-zero value of $\gamma_s$ with the emergence of a plateau. In Fig. \ref{fig5}, we discuss the origin of these characteristics and illustrate further. The parameters for this simulation are set at: $\gamma_f=1,\omega=1,\tau_s=1$.}
    \label{fig1}
\end{figure}
\subsection*{\textit{Underlying reset-free process}}

Combining \eref{i0t-jeff} with \eref{x2t}, one obtains the exact expression for the MSD of the underlying process given by 
\begin{align}
      \langle x^2(t)\rangle&=\frac{k_B T}{m\omega^2}\Bigg[1-e^{-\frac{2\alpha t}{\tau_s}}\cosh^2\left(\sqrt{\alpha^2-\beta^2}t/\tau_s\right)\nonumber\\
&\times\left\{1+\left(\frac{\alpha-\beta^2}{\sqrt{\xi}}\right)\tanh\left(\sqrt{\alpha^2-\beta^2}t/\tau_s\right)\right\}^2\Bigg].
\label{msd-und}
\end{align}
\fref{fig1} depicts the MSD of a particle following GLE for various values of $\gamma_s$. First note that, for $\gamma_s=0$ we obtain
\begin{align}
  \langle x^2 (t) \rangle=  \frac{k_B T}{m \omega^2} \left(1-e^{-\frac{2 t \omega ^2}{\gamma_f}}\right),
\end{align}
which is the MSD of the classical Ornstein-Uhlenbeck (OU) process. The MSD increases linearly, \textit{i.e.}, $\langle x^2 (t) \rangle \sim \frac{2k_BT t}{m \gamma_f} $ at short times (\textit{i.e.} $t\ll \gamma_f/\omega^2$). At large enough times (\textit{i.e.} $t\gg \gamma_f/\omega^2$) the system reaches the \textit{equilibrium stationary state} in a harmonic potential and the MSD saturates to $\langle x^2 (t) \rangle \sim \frac{k_BT }{m\omega^2}$.  

\begin{figure}
    \centering
    \includegraphics[width=8.25cm]{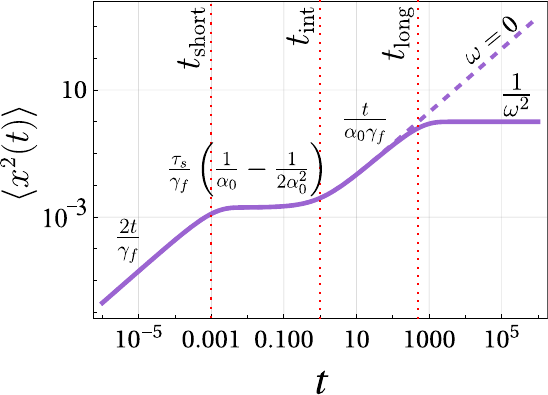}
     \caption{Variation of the MSD of the underlying process (the solid line) as a function of time and emergence of the plateaus for parameters fixed at $\gamma_s=1000,\gamma_f=1,\omega=1,\tau_s=1$. This is the bottom-most curve for the MSD as borrowed from Fig. \ref{fig1}. The vertical dotted lines represents three three different timescales as in \eref{tsshort}, \eref{tsint} and \eref{tslong}. The asymptotic behaviour of the MSD at different timescales is also shown above the curve in each region. The dashed line shows the MSD with $\omega=0$ as in \eref{msd-0}.}
    \label{fig5}
\end{figure}

The memory in the system is introduced through the exponential term in the friction kernel~\eref{jfm-1}, \textit{i.e.}, when $\gamma_s$ is non-zero. From \fref{fig1} note that an intermediate saturation region appears in the MSD due to the effects of memory ($\gamma_s \ne 0$) in GLE compared to the standard Langevin equation ($\gamma_s =0$). The MSD increases linearly, then saturates to an intermediate plateau, then increases again and eventually saturates to the steady state value. In what follows, we find the exact timescales showing the crossover between these distinct regimes and then extract the exact asymptotic behavior of the MSD in all of these regions. Throughout the calculations we shall set $k_B T/m=1$ since it is just a multiplicative factor in the expression for MSD.

The analysis becomes simpler when we set $\omega=0$. The effect of $\omega$ only shows up at very large timescales when the particle starts to feel the effect of the trap and MSD begins to saturate. At short times, $\omega$ does not play any significant role in determining the MSD. Consequently, at short enough time the results with $\omega=0$ and $\omega \ne 0$ match quite well (see \fref{fig5}).

\subsubsection{
\textit{\textbf{The case without potential ($\omega=0$):}}} Taking $\omega \to 0$ is equivalent to $\beta\to 0$ as can be seen from the expression for $\beta=\sqrt{\frac{\omega^2\tau_s}{\gamma_f}}$ (also given in \eref{al-be}) and $\alpha$ becomes
\begin{align}
    \alpha_0=\alpha({\omega \to 0})= \frac{\gamma_f+\gamma_s}{2\gamma_f}.
\end{align}
In this case the MSD takes comparatively simpler form given by
\begin{align}
  \langle x^2 (t) \rangle_{w\to 0}&=  \frac{\tau_s}{\gamma_f}\left(\frac{1}{\alpha_0 }-\frac{1}{2 \alpha_0 ^2}\right)+\frac{t}{\alpha_0  \gamma_f } \nonumber\\
  &\hspace{1cm}-e^{-2 t \alpha_0/\tau_s} \frac{\tau_s}{\gamma_f}\left(\frac{1}{\alpha_0 }-\frac{1}{2 \alpha_0 ^{2}}\right). \label{msd-0}
\end{align}
The above equation has two timescales rooted inside. The short time behavior of this MSD can be obtained when $t\ll \frac{\tau_s}{2\alpha_0}$ and expanding the exponential in the last term. Evidently, the shortest timescale $ t_{\text{short}}$ in the system is given by 
\begin{align}
    t_{\text{short}}=\frac{\tau_s}{2\alpha_0}. \label{tsshort}
\end{align}
In this limit one can find
\begin{align}
\langle x^2 (t ) \rangle_{w\to 0} \approx \frac{2t}{\gamma_f}, ~~ \text{when}~~ t\ll t_{\text{short}}.
\end{align}
Note that any signature of the exponential 
term in the kernel is absent here and the MSD grows linearly with an effective diffusion constant $D=\frac{1}{\gamma_f}$ as is also observed in \fref{fig1}. In contrast, when $t\gg t_{\text{short}}$ one can neglect the exponential term in \eref{msd-0} which yields the following result,
\begin{align}
\langle x^2 (t ) \rangle_{w\to 0} \approx \frac{\tau_s}{\gamma_f}\left(\frac{1}{\alpha_0 }-\frac{1}{2 \alpha_0 ^2}\right)+\frac{t}{\alpha_0  \gamma_f },~\text{when}~t\gg t_{\text{short}}.
\end{align}
The above result further shows distinct behaviors below and above an intermediate timescale $t_{\text{int}}$. To find this intermediate timescale let us rewrite the expression in the following way
\begin{align}
\langle x^2 (t ) \rangle_{w\to 0} &\approx \frac{\tau_s}{\gamma_f}\left(\frac{1}{\alpha_0 }-\frac{1}{2 \alpha_0 ^2}\right)\left[1+\frac{t}{ \tau_s \left(1-\frac{1}{2 \alpha_0}\right) }\right]\nonumber\\
&\approx \frac{\tau_s}{\gamma_f}\left(\frac{1}{\alpha_0 }-\frac{1}{2 \alpha_0 ^2}\right)\left[1+\frac{t}{t_{\text{int}}}\right]. \label{msd-int}
\end{align}
Evidently, the intermediate timescale is given by
\begin{align}
    t_{\text{int}}=\tau_s \left(1-\frac{1}{2 \alpha_0}\right). \label{tsint}
\end{align}
When $t\ll t_{\text{int}} $ the second time dependent term in \eref{msd-int} can be neglected and the MSD attains saturation value 
\begin{align}
   \langle x^2 (t ) \rangle_{w\to 0} \approx \frac{\tau_s}{\gamma_f}\left(\frac{1}{\alpha_0 }-\frac{1}{2 \alpha_0 ^2}\right), ~ \text{when}~t_{\text{short}} \ll t\ll t_{\text{int}}.
\end{align}
The above result gives analytical expression for the first plateau as is seen in \fref{fig1}. On the other hand when $t\gg t_{\text{int}}$ we can neglect the constant value of unity in the third parenthesis of \eref{msd-int} to have
\begin{align}
   \langle x^2 (t ) \rangle_{w\to 0} \approx\frac{t}{\alpha_0  \gamma_f }, ~~ \text{when}~~t\gg t_{\text{int}}.
\end{align}
Thus the MSD grows linearly after the intermediate region ends as also seen in \fref{fig1}.
Let us now examine the case with $\omega \ne 0$.

\subsubsection{
\textit{\textbf{The case with potential ($\omega\ne 0$):}}} To find the longest timescale $t_{\text{long}}$   we first note that the MSD saturates to the \textit{equilibrium} steady state value  $1/\omega^2$ at long enough times $t\gg t_{\text{long}}$.  This result can be verified by taking the limit $t\to \infty$ in the expression for MSD in \eref{msd-und}. However, one can also find the correction to this term in the limit $t\to \infty$, the approximate expression for which is found to be
\begin{align}
     \langle x^2 (t \to \infty) \rangle \approx \frac{1}{\omega^2}-A e^{-\frac{2 t \left(\alpha -\sqrt{\alpha^2-\beta^2} \right)}{\tau_s }},
\end{align}
where $A$ is a time-independent prefactor dependent on $\alpha,~ \beta, ~\tau_s$ and $\omega$. The above asymptotic expression for the MSD immediately reveals the longest timescale $t_{\text{long}}$ of the system which is given by
\begin{align}
    t_{\text{long}}=\frac{\tau_s}{2\left(\alpha- \sqrt{\alpha^2-\beta^2}\right)}. \label{tslong}
\end{align}
Beyond this timescale the MSD saturates to the steady state value. Finally combining all the above results we have the MSD of the underlying system at different timescales given by (with $\frac{k_B T}{m}=1$),
\begin{eqnarray}
\langle x^2 (t ) \rangle\approx
\begin{cases}
 \frac{2t}{\gamma_f} &\text{when }~ t\ll t_{\text{short}}, \vspace{0.1cm}\\ 
\frac{\tau_s}{\gamma_f}\left(\frac{1}{\alpha_0 }-\frac{1}{2 \alpha_0 ^2}\right)  &\text{when }~ t_{\text{short}} \ll t\ll t_{\text{int}},~ \\
\frac{t}{\alpha_0  \gamma_f } & \text{when}~~t_{\text{long}} \gg t\gg t_{\text{int}},\vspace{0.1cm}\\

\frac{1}{\omega^2}& \text{when}~~t\gg t_{\text{long}}.
\end{cases}
\label{msd-asmp}
\end{eqnarray}

To illustrate these timescales, we select the bottom most plot in \fref{fig1} and plot it separately in \fref{fig5}. Here, the parameters are fixed at $\gamma_s=1000,\gamma_f=1,\omega=1,\tau_s=1$, so that $\alpha=501, \alpha_0=500.5, \beta=1$. In \fref{fig5} we show the different timescales of the system and the asymptotic behavior of the MSD in each of this region. Note that the plot with $\omega=0$ shown by the dashed line mathces exactly with the plot for $\omega \ne 0$ at $t \ll t_{\text{long}}$ which validates our earlier assumption.

\begin{figure}
    \centering
    \includegraphics[width=8.4cm]{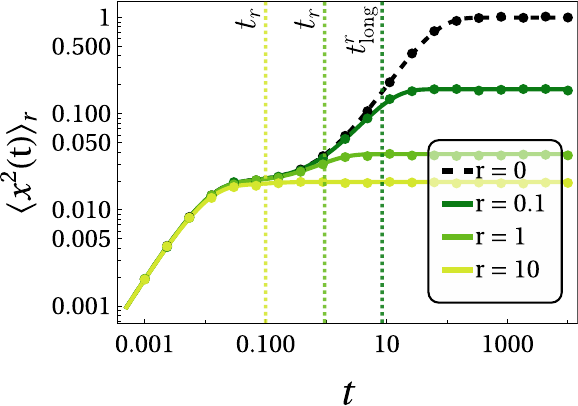}
    \caption{Variation of MSD with time for the resetting system for different values of resetting rate $r$. The parameters are fixed at: $\gamma_f=1,\gamma_s=100,\tau_s=1,\omega=1$. The circles represent results from the simulation. The dashed curve represents the result for the underlying reset-free process, \textit{i.e.}, $r=0$. The solid circles represent the results from numerical simulation. The rightmost vertical dashed line represents the longest timescale of the system under resetting while the other two  vertical dashed lines are associated with the timescales $t_r$ for different values of $r$ (color-codes represent the values of $r$ respectively). At sufficiently short times $t \ll t_r, t_{\text{long}}^r$, the curves follow the MSD of the underlying reset-free process.}
    \label{fig2}
\end{figure}
\subsection*{\textit{Resetting induced process}}

Let us now delve into the details of MSD under stochastic resetting. Plugging \eref{i0t-jeff} into \eref{msd-r} gives the exact expression for the MSD under resetting. The expression is quite lengthy and not very insightful so we have moved that to the \aref{appa} (in particular Eq. (\ref{msd-jeff}1)).  In \fref{fig2} we show its behaviour with respect to time. The dashed line in \fref{fig2} represents the MSD without resetting. First note that resetting introduces another extra timescale
\begin{align}
    t_r=1/r,
\end{align}
 in the system which is the average waiting time between two resetting intervals. 
Note that this timescale is independent of the system and entirely controlled externally. From \fref{fig2} it is evident that the effects of resetting show up only at the longest timescale of the system. After expanding the expression for MSD with resetting as given in Eq. (\ref{msd-jeff}1) in the limit $t\to \infty$ one finds
\begin{align}
      \langle x^2 (t \to \infty) \rangle_r \approx   \langle x^2  \rangle_r^{ss}- B e^{-\frac{t \left(2 \alpha-2 \sqrt{\alpha^2-\beta^2}+ r \tau_s \right)}{\tau_s }}, \label{msdr-asym}
\end{align}
where $\langle x^2  \rangle_r^{ss}$ is the steady state MSD under resetting given by 
\begin{align}
\langle x^2 \rangle_r^{ss}= \frac{2 \beta ^2 \left[4 \alpha  \left(r \tau _s+1\right)+r \tau _s \left(1+r \tau _s-\beta ^2\right)\right]}{\omega ^2 \left(2 \alpha +r \tau _s\right) \left[4 \beta ^2+r \tau _s \left(4 \alpha +r \tau _s\right)\right]},
 \label{eq:x2ss}
\end{align}
and $B$ is just a time-independent constant.  As a consistency check note that in the limit $r\to 0$, one finds $\langle x^2 \rangle_{r\to 0}^{ss}=1/\omega^2$ which was also obtained earlier. It should be noted that the steady state under resetting is not a pure equilibrium as the probability current due to resetting continuously flows through the system even at large time. Thus this saturation value of the MSD is qualitatively distinct from the same under equilibrium condition imposed by the harmonic trap. From \eref{msdr-asym} one can infer the longest timescale of the system under resetting $t_{\text{long}}^r$ given as
\begin{align}
    t_{\text{long}}^r=\frac{\tau_s}{2\left(\alpha- \sqrt{\alpha^2 -\beta^2  }\right)+ r \tau_s}. \label{tlongr}
\end{align}
When the resetting rate is very low so that $\frac{1}{r} \gg \frac{\tau_s}{2\left(\alpha- \sqrt{\alpha^2-\beta^2}\right)} \quad \text{or} \quad t_r \gg t_{\text{long}}$, the term $r \tau_s$ in the denominator of \eref{tlongr} can be neglected and we have $t_{\text{long}}^r \approx t_{\text{long}}$. Thus the steady state with or without resetting is obtained at the same timescale, although the exact steady state value of the MSD in both cases are different.

In turn, for a suitably finite value of resetting rate $r$ one has $t_{\text{long}}^r < t_{\text{long}}$. Hence the steady state with resetting occurs at at a time earlier than the longest timescale of the underlying process. Note that at very short times $t \ll  t_{\text{long}}^r$ the effect of resetting does not show up in the system the MSD merges that with that of the underlying process as can be seen in \fref{fig2}. When the resetting rate is too high one can neglect the first term in the denominator of \eref{tlongr} that results in $ t_{\text{long}}^r \approx  t_{r}=1/r$. Steady state is obtained at times $t \gg t_r$ as evident from \fref{fig2}. As effect of resetting is absent below $t_r$ thus all the other timescales of the underlying process remains intact (plot with $r=0.1$ in \fref{fig2}).

In the limit of significantly high resetting rate so that $t_r$ is less than either  $t_{\text{int}}$ or $t_{\text{short}}$, then  $t_{r}$ remains the dominant timescale and the other ones vanishes. As an illustrative example consider the curve in \eref{fig2} with $r=10$. Here the resetting timescale $t_r=0.1$ is shorter than the intermediate timescale $t_{\text{int}}=1$. As a result, the intermediate timescale does not show up in the MSD further causes the intermediate plateau to diminish. Here, the steady state is obtained just after $t_r=0.1$ which is much earlier than $t_{\text{int}}$. For sufficiently high resetting rate thus one obtains the following value for the steady state MSD
\begin{align}
    \langle x^2 \rangle_{r \to \infty}=\frac{1}{\gamma_f r},
\end{align}
which is the same for a free Brownian particle with diffusion constant $D=\frac{1}{ \gamma_f}$ \cite{evans2011diffusion}.

\subsection{Correlation function} \label{secIVB}
\begin{figure}
    \centering
    \includegraphics[width=8.4cm]{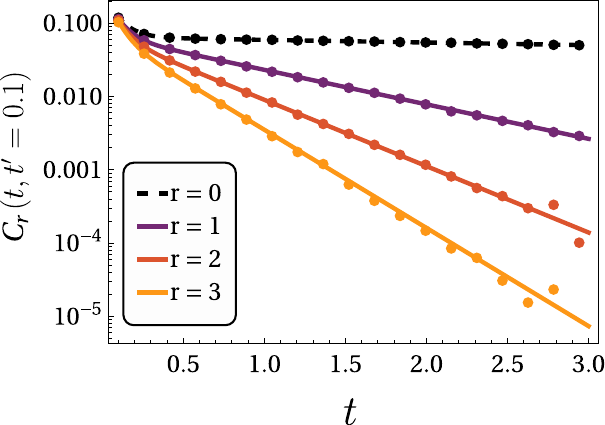}
    \caption{Variation of the correlation function with time under resetting for Jeffreys fluid model. The parameters chosen are: $\gamma_f=1, \gamma_s=10, \omega=1, \tau_s=1$. As before, the dashed line represents the reset-free process ($r=0$) and the solid circles represent data points from numerical simulation. Note that the correlation decays exponentially with time as given by \eref{corr-asym}. }
    \label{fig3}
\end{figure}

We now turn our attention to the analysis of the correlation function in the Jeffreys fluid model. Plugging \eref{i0t-jeff} in \eref{xclt-ovd} gives us the correlation function for the underlying reset-free process. At large enough time, the underlying process's correlation function decays exponentially with time $t$ (keeping $t'$ fixed) as
\begin{align}
    C(t,t';~t\gg t') \sim e^{-\frac{ \left(\alpha -\sqrt{\alpha ^2-\beta ^2}\right)t}{\tau_s }}, \label{corr-und}
\end{align}
where recall that $\alpha=\frac{\gamma_f+\gamma_s+\tau_s\omega^2}{2\gamma_f}$ and $\beta=\sqrt{\frac{\omega^2\tau_s}{\gamma_f}}$. The correlation is expected to die off as the system equilibrates. However, note that for $\omega \to 0$, the correlation function saturates to a fixed value indicating that there is no steady state as can be corroborated from \fref{fig5} which displays the linear growth of MSD in time. The correlation function under resetting interrupted dynamics can be found by inserting \eref{xclt-ovd} in \eref{corr-reset} with $I_0(t)$ given by \eref{i0t-jeff}. The exact expression for the correlation function is provided in \aref{appa}. \fref{fig3} shows the behavior of the correlation function with respect to time $t$ keeping $t'=0.5$ fixed. Under resetting the correlation function decays exponentially, faster than the underlying process, with the following asymptotic behavior (see \aref{appa} for more details)
\begin{align}
     C_r(t,t';~t\gg t') \sim e^{-rt-\frac{ \left(\alpha -\sqrt{\alpha ^2-\beta ^2}\right)t}{\tau_s } }. \label{corr-asym}
\end{align}
Notably, the correlation function under resetting vanishes in the long time as we take the $\omega \to 0$ limit. This is a manifestation of the system reaching the steady state under resetting dynamics which was not the case for the underlying process even when the trap is turned off. Thus, resetting plays a crucial role in stabilizing the system especially in the absence of the potential.

\section{Simulation scheme for GLE with resetting} \label{sec5}
So far, we have focused on deriving exact results for the MSD and the correlation functions. The aim of this section is to provide the steps to solve the GLE-systems numerically. Unlike the classical Langevin equation (given by \eref{le}), solving the GLE (given by \eref{GLE}) numerically can be challenging at times as can be perceived from the earlier literatures where 
various simulation schemes have been proposed to handle such systems \cite{duong2022accurate,wisniewski2024effective,bockius2021model,baczewski2013numerical}. In this work we follow the Markovian embedding method which has been successful in implementing non-Markovian systems especially the GLEs that represent colloids in viscoelastic medium \cite{baczewski2013numerical,siegle2010markovian,siegle2011markovian,das2023enhanced}.

It will be useful to recall the 
the overdamped GLE for the Jeffreys fluid model as in \eref{jfm-1} namely
\begin{align}
    \gamma_f \dot{x}(t)+\frac{\gamma_s}{\tau_s}\int_{0}^t \exp\left(-\frac{t-t'}{\tau_s}\right)\dot{x}(t')dt' +\omega^2 x(t)=\xi (t), \label{gle-2}
\end{align}
where we have set $m=1$. The above non-Markovian equation is difficult to solve numerically for the following reasons:
\begin{itemize}
    \item The second term requires storage of the velocity for all the time prior to $t$ which is computationally expensive. In addition, to find the time dependence of the statistical quantities, at each measurement time step one needs to calculate the convolution of the kernel as well as the velocity upto the measurement time which costs a significant computational time. 
    \item Generating the noise $\xi (t)$ which has the correlation as in \eref{fdt} requires storage of correlated random numbers up to time $t$. For instance, to generate a single trajectory upto time $t$ with a microscopic time interval $\Delta t=10^{-4}$, one needs to keep track of the noise history upto time $t$ unlike the Markovian process where noise acts independently at each time step. So, to gather the value of MSD, say at time $t=10^4$, one should store around  $\sim 10^8$ data points of $\xi(t)$ (of approximate file size $\sim 400$ MB). Now, for a better averaging with a sample set of $N$ trajectories, one need to generate an equal set of random numbers  each containing $10^8$ data points of $\xi(t)$ while running every trajectory. This task turns out to be enormously time consuming. 
\end{itemize}
To circumvent these numerical hindrances, we use the Markovian embedding method which turns out to be more efficient. Below, we briefly describe this method and underpin how this evades both the problems mentioned in above.
\subsection*{\textbf{\textit{Markovian embedding}}}
To circumvent the first problem we define an auxiliary variable $W(t)$ as
\begin{align}
    W(t)=\frac{\gamma_s}{\tau_s}\int_{0}^t \exp\left(-\frac{t-t'}{\tau_s}\right)\dot{x}(t')dt'.
\end{align}
Taking derivative with respect to $t$ of the above equation yields
\begin{align}
    \dot{W}(t)=-\frac{1}{\tau_s}W(t)+\frac{\gamma_s}{\tau_s}\dot{x}(t).
\end{align}
This step reduces the complexity quite a bit since the single non-Markovian equation \eref{gle-2} now reduces to a set of two Markovian equations as we note below
\begin{align}
    &\dot{x}=-\frac{1}{\gamma_f}W(t) -\frac{\omega^2}{\gamma_f}x(t) +\frac{1}{\gamma_f}\xi(t), \nonumber\\
    &\dot{W}(t)=-\frac{1}{\tau_s}\left(1+\frac{\gamma_s}{\gamma_f}\right)W(t)-\frac{\omega^2}{\tau_s}\frac{\gamma_s}{\gamma_f}x(t)+\frac{1}{\tau_s}\frac{\gamma_s}{\gamma_f}\xi(t).
\end{align}
To bypass the second problem related to the correlated noise, we define the OU process $\eta(t)$ defined as
\begin{align}
    \dot{\eta}(t)=-\frac{1}{\tau_s}\eta(t) +\frac{1}{\tau_s} \zeta(t),
\end{align}
where $\zeta(t)$ is a white noise with zero mean and correlation $\langle\zeta(t)\zeta(t')\rangle=2\gamma_s k_B T \delta(t-t')$. Note that the correlation for an OU process $\eta(t)$ is given by
\begin{align}
    \langle \eta(t) \eta(t') \rangle=k_B T \frac{\gamma_s}{\tau_s} \exp\left(\frac{|t-t'|}{\tau_s}\right).
\end{align}
Thus, the OU process $\eta(t)$ is able to generate the exponential correlation part as in \eref{jfm-1}. We write the noise $\xi(t)$ as
\begin{align}
    \xi(t)=\zeta_0(t) +\eta (t),
\end{align}
where $\zeta_0(t)$ is another white noise with correlation $\langle\zeta_0(t)\zeta_0(t')\rangle=2\gamma_f k_B T \delta(t-t')$. Henceforth, we arrive at a situation where one needs to solve the following set of Markovian equations to find $x(t)$ namely
\begin{align}
     &\dot{x}(t)=-\frac{1}{\tau_s}W(t) -\frac{\omega^2}{\gamma_f}x(t) +\frac{1}{\gamma_f}\zeta_0(t)+\frac{1}{\gamma_f}\eta(t), \nonumber\\
    &\dot{W}(t)=-\frac{1}{\tau_s}\left(1+\frac{\gamma_s}{\gamma_f}\right)W(t)-\frac{\omega^2}{\tau_s}\frac{\gamma_s}{\gamma_f}x(t)\nonumber\\
    &\hspace{4cm}+\frac{1}{\tau_s}\frac{\gamma_s}{\gamma_f}\zeta_0(t)+\frac{1}{\tau_s}\frac{\gamma_s}{\gamma_f}\eta(t),\nonumber\\
    &\dot{\eta}(t)=-\frac{1}{\tau_s}\eta(t) +\frac{1}{\tau_s} \zeta(t), \label{gle-me}
\end{align}
with the correlation of the white noises $\zeta(t),\zeta_0(t)$ given by
\begin{align}
    &\langle\zeta(t)\zeta(t')\rangle=2\gamma_s k_B T \delta(t-t'),\\
    &\langle\zeta_0(t)\zeta_0(t')\rangle=2\gamma_f k_B T \delta(t-t').
\end{align}
We mention in passing the initial condition $\overrightarrow{X}(0)=(x(0),W(0),\eta(0))\equiv (0,0,\eta_0)$ where $\eta_0$ is a randomly chosen number from the normal distribution with zero mean and variance $\sqrt{\gamma/ \tau_s}$. 

\subsection*{\textbf{\textit{Stochastic resetting}}}
Note that, in our theory, resetting the particle refers to resetting both its position and the friction kernel of the viscoelastic medium back to their initial value. Consequently, to implement a resetting event, we set the particle's position along with the auxiliary variables back to the starting configuration, \textit{i.e.}, $\overrightarrow{X}(t)=(x(t),W(t),\eta(t))=\overrightarrow{X}(0)\equiv (0,0,\eta_0)$ at the resetting times.  As each resetting event occurs at an exponentially distributed times with rate $r$ the particle's motion at each time step $\Delta t$ is governed by the following rules:
\begin{eqnarray}
\overrightarrow{X}(t+ \Delta t)=
\begin{cases}
 \overrightarrow{X}(0), &\text{w.p. }\; r\Delta t, \\ \\
\text{follows \eref{gle-me}},  &\text{w.p.}\;\; (1-r\Delta t), 
\end{cases}
\label{lg}
\end{eqnarray}
with w.p. implying \textit{with probability}. For the simulation, we chose $\Delta t = 10^{-4}$ and averaged the data over $\approx 10^4$ trajectories for the MSD and $\approx 10^6$ trajectories for the correlation function, respectively.

\section{Discussion and outlook} \label{secVI}
Over the recent years, stochastic resetting has gained considerable attention in the field of statistical physics, stochastic process, chemical \& biological process and in many interdisciplinary studies. The studies are made to its very depth both in theory \cite{evans2011diffusion,pal2015diffusion,gupta2014fluctuating,sokolov2023linear} and in experiments \cite{tal2020experimental,besga2020optimal,paramanick2024uncovering}. We refer to \cite{evans2020stochastic,pal2023random,pal2022inspection,gupta2022stochastic} for a comprehensive review of this subject.
In parallel, numerous studies have successfully applied GLE with different kernels to understand motion of bio-molecules in crowded environment or colloids in elastic medium
\cite{lange2006collective,lee2019multi,goychuk2012viscoelastic}. This work is a first step to merge both the fields and unravel the statistical properties of the GLE under stochastic resetting with an attempt that it will render useful insights similar to the paradigm of diffusion under resetting. 

To begin with, 
we extended the standard renewal formalism to accommodate Langevin systems with memory. This allowed us to derive exact expressions for the MSD and correlation function of a resetting particle in a viscoelastic medium with arbitrary friction kernel. These generalized results are illustrated for the Jeffreys fluid model. While the MSD without resetting under Jeffreys fluid model shows an intermediate saturation region, which is a fingerprint of the slow relaxation timescale of the viscoelastic bath, our work shows that resetting can harness this saturation region, and for a very high resetting rate the particle is seen to effectively follow the classical diffusive behavior. In addition, we find that the correlation function decays faster in the presence of resetting. We corroborate our theoretical findings with numerical simulations of the GLE under resetting, using the Markovian embedding approach.

We believe that this work will open several future research avenues. One immediate next step would be to treat the same problem for a different kernel (e.g. power law which is ubiquitous in anomalous diffusion phenomena \cite{min2005observation,min2006kramers,sandev2015diffusion,slkezak2018superstatistical}) and see the effect of resetting pertaining to the steady state properties. Moreover, there has been substantial advancement in recent time going beyond Markovian resetting strategies \cite{pal2016diffusion,pal2017first,radice2022diffusion}, and with the inclusion of practically feasible space-time coupled resetting \cite{bodrova2020resetting,tucci2022first,biswas2024search,pal2020search,gupta2020stochastic} -- understanding these ramifications on the current system of interest would be the next promising step. Finally, recent experiments using optical traps and colloids in viscoelastic medium have reported ample interesting properties related to the transport properties or stochastic energetics \cite{paul2021bayesian,das2023enhanced,gomez2021work,darabi2023stochastic}. It will be worthwhile to verify some of our results under resetting dynamics using these precision experiments.

\section{Acknowledgment}
We are thankful to Biswajit Das for fruitful discussion and providing useful references on the methods of simulations for solving GLE. The numerical calculations reported in
this work were carried out on the Nandadevi and Kamet cluster, which are maintained and supported by the Institute of Mathematical Science’s High-Performance Computing Center. AP gratefully acknowledges research support from the Department of Atomic Energy, Government of India via Soft Matter Apex projects. TS acknowledges financial support by the German Science Foundation (DFG, Grant number ME 1535/12-1) and by the Alliance of International Science Organizations (Project No. ANSO-CR-PP-2022-05). TS was also supported by the Alexander von Humboldt Foundation.
\vspace{0.2cm}
\appendix
\section{Exact expressions for the MSD and correlation function}
\label{appa}
In this section, we provide the exact analytical expression for the MSD with resetting, \textit{i.e.}, $\langle x^2 (t) \rangle_r$ (plotted in \fref{fig2}) and the correlation function both with and without resetting, \textit{i.e.}, $C_r(t,t')$ (plotted in \fref{fig3}) respectively, for Jeffreys fluid model. In both the expressions we assume $k_BT/m=1$. With $\alpha,\beta$ defined as in \eref{al-be}, the MSD with resetting is given by
\begin{widetext}
\begin{small}
    \begin{align}
     \langle x^2 (t) \rangle_r =&\frac{\beta^2}{\omega ^2 (\alpha^2-\beta^2) \left(4 \beta ^2+r \tau _s \left(4 \alpha +r \tau _s\right)\right)}\Bigg[ e^{-t \left(r+\frac{2 \alpha }{\tau _s}\right)} \Bigg(\sqrt{\alpha^2-\beta^2} \left(-4 \alpha +4 \beta ^2+\left(\beta ^2-1\right) r \tau _s\right) \sinh \left(\frac{2 t \sqrt{\alpha ^2-\beta ^2}}{\tau _s}\right) \nonumber\\
      & \hspace{4.5cm} -\left(4 \alpha ^2-2 (2 \alpha +1) \beta ^2+2 \beta ^4+r \left((\alpha -2) \beta ^2+\alpha \right) \tau _s\right) \cosh \left(\frac{2 t \sqrt{\alpha ^2-\beta ^2}}{\tau _s}\right)\Bigg) \nonumber\\
     &  \hspace{1.3cm}+ \frac{2 (\alpha^2-\beta^2) \left(4 \alpha +r \tau _s \left(4 \alpha -\beta ^2+r \tau _s+1\right)\right)-\alpha  \left(2 \alpha -\beta ^2-1\right) \left(4 \beta ^2+r \tau _s \left(4 \alpha +r \tau _s\right)\right) e^{-t \left(r+\frac{2 \alpha }{\tau _s}\right)}}{2 \alpha +r \tau _s}\Bigg].
    \end{align}
    \end{small}\label{msd-jeff}
\end{widetext}
The correlation function of the underlying process \textit{i.e.} $C(t,t')$ for $t>t'$ is given by

\begin{widetext}
    \begin{align}
        &C(t,t')=\frac{e^{-\frac{\alpha  (t'+t)}{\tau _s}}}{2 \omega ^2 (\alpha^2 -\beta^2)}\vast(\left(\beta ^2 \left(-2 \alpha +\beta ^2+1\right)+2 \alpha^2 -\beta^2 e^{\frac{2 \alpha  t'}{\tau _s}}\right) \cosh \left(\frac{\sqrt{\alpha^2 -\beta^2 } (t-t')}{\tau _s}\right)\nonumber\\
        &\hspace{7cm}+\left(-2 \alpha ^2+2 \alpha  \beta ^2-\beta ^4+\beta ^2\right) \cosh \left(\frac{\sqrt{\alpha^2 -\beta^2 } (t'+t)}{\tau _s}\right) \nonumber\\
        &\hspace{4cm}-2 \sqrt{\alpha^2 -\beta^2 } \left(\alpha -\beta ^2\right) \left(\sinh \left(\frac{\sqrt{\alpha^2 -\beta^2 } (t'+t)}{\tau _s}\right)-e^{\frac{2 \alpha  t'}{\tau _s}} \sinh \left(\frac{\sqrt{\alpha^2 -\beta^2 } (t-t')}{\tau _s}\right)\right)\vast).
    \end{align}
\end{widetext}

The correlation function under resetting, after utilizing \eref{xclt-ovd} and \eref{corr-reset}, is given by 
\pagebreak
\begin{widetext}
    \begin{align}
     &C_r(t,t') = \frac{e^{-r (t-t')}}{2 \omega ^2 (\beta^2 -\alpha^2 ) \left(4 \beta ^2+r \tau _s \left(4 \alpha +r \tau _s\right)\right)} \Vast(\frac{r \tau _s e^{-r t'-\frac{\alpha  (t'+t)}{\tau _s}} \cosh \left(\frac{-\sqrt{\alpha^2-\beta^2} (t-t')}{\tau _s}\right)}{2 \alpha +r \tau _s} \times \nonumber\\
     & \Big[\beta ^2 \left(-2 \alpha +\beta ^2+1\right) \left(4 \beta ^2+r \tau _s \left(4 \alpha +r \tau _s\right)\right)+2 (\alpha^2-\beta^2) \left(2 \left(\left(\beta ^2-2 \alpha \right)^2+\beta ^2\right)+r \tau _s \left(6 \alpha -2 \beta ^2+r \tau _s\right)\right) e^{t' \left(r+\frac{2 \alpha }{\tau _s}\right)}\Big] \nonumber\\
     &+e^{-3 r t'-\frac{\alpha  (3 t'+t)}{\tau _s}} \vast[-e^{2 t' \left(r+\frac{\alpha }{\tau _s}\right)} \left(\beta ^2 \left(-2 \alpha +\beta ^2+1\right)+2 (\alpha^2-\beta^2) e^{t' \left(r+\frac{2 \alpha }{\tau _s}\right)}\right) \cosh \left(\frac{\sqrt{\alpha^2-\beta^2} (t-t')}{\tau _s}\right) \times \nonumber\\
     &(4 \beta ^2+r \tau _s \left(4 \alpha +r \tau _s\right))+2 \beta^2\Bigg[e^{2 t' \left(r+\frac{\alpha }{\tau _s}\right)} \cosh \left(\frac{\sqrt{\alpha^2-\beta^2} (t'+t)}{\tau _s}\right)(4 \alpha ^2-2 (2 \alpha +1) \beta ^2+2 \beta ^4+r \left((\alpha -2) \beta ^2+\alpha \right) \tau _s) \nonumber\\
     &\hspace{4cm}-\sqrt{\alpha^2-\beta^2}(4 \alpha -\left(\beta ^2 \left(r \tau _s+4\right)\right)+r \tau _s)\Big[e^{3 r t'+\frac{4 \alpha  t'}{\tau _s}} \sinh \left(\frac{\sqrt{\alpha^2-\beta^2} (t-t')}{\tau _s}\right) \nonumber\\
     &\hspace{11cm}-e^{2 t' \left(r+\frac{\alpha }{\tau _s}\right)} \sinh \left(\frac{\sqrt{\alpha^2-\beta^2} (t'+t)}{\tau _s}\right)\Big]\Bigg]\vast]\Vast).
    \end{align}
\end{widetext}

\subsection*{\textbf{\textit{Asymptotic behaviour of the correlation function}}}
Let us now provide the exact asymptotic forms of the correlation function which was discussed briefly in section \ref{secIVB}. The exact asymptotic form of the correlation function of the underlying reset-free process is given by
\begin{widetext}
    \begin{align}
         & C(t,t';~t\gg t') \sim \frac{\exp \left(-\frac{t' \left(\sqrt{\alpha^2 -\beta^2}+\alpha \right)+t \left(\alpha -\sqrt{\alpha^2 -\beta^2}\right)}{\tau _s}\right)}{4 \omega ^2 (\alpha^2 -\beta^2 )}\vast(2 \left(\alpha ^2-\beta ^2 \left(\sqrt{\alpha^2 -\beta^2}+1\right)+\alpha  \sqrt{\alpha^2 -\beta^2}\right) e^{\frac{2 \alpha  t'}{\tau _s}}\nonumber\\
          &\hspace{2cm}+\beta ^2 \left(-2 \alpha +\beta ^2+1\right)+e^{\frac{2 t' \sqrt{\alpha ^2-\beta ^2}}{\tau _s}}\left(-2 \alpha ^2+2 \alpha  \left(\beta ^2-\sqrt{\alpha^2 -\beta^2}\right)+\beta ^2 \left(2 \sqrt{\alpha^2 -\beta^2}-\beta ^2+1\right)\right)\vast).
    \end{align}
\end{widetext}
Keeping $t'$ fixed, in the limit $t\to \infty$ note that the term under the parenthesis is just a constant prefactor with the only $t$ dependence coming through the exponential term outside which is $e^{-\frac{ \left(\alpha -\sqrt{\alpha ^2-\beta ^2}\right)t}{\tau_s }}$, the result provided in \eref{corr-und}.  

On the other hand, the asymptotic behavior of the correlation function under resetting is given by
\begin{widetext}
    \begin{align}
        &C_r(t,t';~t\gg t') \sim \frac{\beta ^2 \exp \left(-\frac{t \left(-\sqrt{\alpha^2 -\beta^2}+\alpha +r \tau_s \right)+t' \left(\sqrt{\alpha^2 -\beta^2}+\alpha \right)}{\tau_s }\right)}{2 \omega ^2 (\alpha^2 -\beta^2) (2 \alpha +r \tau_s ) \left(4 \beta ^2+r \tau_s  (4 \alpha +r \tau_s )\right)} \vast(\alpha  \left(-2 \alpha +\beta ^2+1\right) \left(4 \beta ^2+r \tau_s  (4 \alpha +r \tau_s )\right) \nonumber\\
        &\hspace{2cm}(2 \alpha +r \tau_s ) e^{\frac{2 t' \sqrt{\alpha ^2-\beta ^2}}{\tau_s }}\Bigg[4 \alpha ^2+2 \beta ^4+r \tau_s  \sqrt{\alpha^2 -\beta^2}-\beta ^2 \left(4 \sqrt{\alpha^2 -\beta^2}+r \tau_s  \left(\sqrt{\alpha^2 -\beta^2}+2\right)+2\right) \nonumber\\
        &\hspace{1cm}+\alpha  \left(4 \sqrt{\alpha^2 -\beta^2}+\beta ^2 (r \tau_s -4)+r \tau_s \right)\Bigg]+ e^{t' \left(\frac{2 \alpha }{\tau_s }+r\right)} \Bigg[8 \alpha ^3 (r \tau_s +1)+2 \alpha ^2 \left(r \tau_s  \left(-\beta ^2+r \tau_s +1\right)-4 \sqrt{\alpha ^2-\beta ^2}\right) \nonumber\\
        &\hspace{6cm}+2 \alpha  \left(\beta ^2 \left(4 \left(\sqrt{\alpha^2 -\beta^2}-1\right)+r \tau_s  \left(\sqrt{\alpha^2 -\beta^2}-4\right)\right)-3 r \tau_s  \sqrt{\alpha^2 -\beta^2}\right) \nonumber\\
        &\hspace{5cm}+r \tau_s  \left(2 \beta ^4+\beta ^2 \left(4 \sqrt{\alpha^2 -\beta^2}+r \tau_s  \left(\sqrt{\alpha^2 -\beta^2}-2\right)-2\right)-r \tau_s  \sqrt{\alpha^2 -\beta^2}\right)\Bigg]\vast).
    \end{align}
\end{widetext}
For fixed $t'$ one can again see that the only $t$ dependent term lies outside the parenthesis and is given by $e^{-rt-\frac{ \left(\alpha -\sqrt{\alpha ^2-\beta ^2}\right)t}{\tau_s } }$, which is \eref{corr-asym} from the main text.

\bibliography{apssamp}

\end{document}